\documentclass[12pt]{article}
\usepackage[latin1]{inputenc}
\usepackage[centertags]{amsmath}
\usepackage{amsfonts}
\usepackage{amssymb}
\usepackage{amsthm}
\usepackage{newlfont}
\usepackage{epsf}
\usepackage{graphicx}
\usepackage[sort&compress,round,comma]{natbib}
\usepackage{geometry}

\newlength{\defbaselineskip}
\newcommand{\setlinespacing}[1]%
           {\setlength{\baselineskip}{#1 \defbaselineskip}}

\newcommand{\singlespacing}{\setlength{\baselineskip}{\defbaselineskip}}

\theoremstyle{plain}
\newtheorem{thm}{Theorem}

\newtheorem{lem}[thm]{Lemma}

\theoremstyle{definition}
\newtheorem{defn}[thm]{Definition}
\newtheorem{axiom}{Axiom}
\theoremstyle{remark}

\newcounter{axiomcounter}
\numberwithin{axiom}{section} 

\def\BL{\mathcal L}

\long\def\symbolfootnote[#1]#2{\begingroup%
\def\thefootnote{\fnsymbol{footnote}}\footnote[#1]{#2}\endgroup}

\begin{document}


\begin{center}
{\LARGE\textbf{Information-theoretic principle entails
orthomodularity of a lattice}}
      \vskip 0.3 true in
{\large \bf Alexei Grinbaum}
\vskip 0.15 true in
\par
{\it CREA, Ecole Polytechnique, 1 rue Descartes 75005 Paris, France
\par Email alexei.grinbaum@polytechnique.edu}
\par\vskip 0.3 true in
\end{center}

\bigskip

\begin{quote}
Quantum logical axiomatic systems for quantum theory usually
include a postulate that a lattice under consideration is
orthomodular. We propose a derivation of orthomodularity from an
information-theoretic axiom. This provides conceptual clarity and
removes a long-standing puzzle about the meaning of
orthomodularity.
\end{quote}
\vskip 0.15 true in
\par

\section{INTRODUCTION}

There exist many axiomatic systems from which one derives the
formalism of quantum
theory.\cite{mackey63,zieler,varada2,piron76,kochspeck65,guenin,Gunson,jauch,pool,plymen,maczy,marlow,holland,ruttimann}
Some of these systems employ the concepts of orthodox quantum logic,
i.e. of the theory of orthomodular lattices. For reconstruction of
quantum theory, orthomodularity of the lattice is a necessary,
although not a sufficient condition.\cite{keller} In the
axiomatizations it is commonly postulated or derived from
\textit{formal} assumptions about relations between elements of the
lattice. The work of Beltrametti and Cassinelli is a notable
exception: they give a {\it conceptual} justification of
orthomodularity. Namely, they argue that ``orthomodularity
corresponds to the survival\ldots\,of a notion of the logical
conditional, which takes the place of the classical implication
associated with Boolean algebra''.\cite{bc} A similar motivation was
chosen by Jauch and Piron, who reformulated orthomodularity as a
condition that if a proposition is greater than another one in the
lattice, then they must be \textit{compatible}, i.e. their span with
the operations of join and meet and the orthogonal complement must
give a Boolean algebra.\cite{piron,jauch} Although his own
motivation of orthomodularity is different, Drieschner gives a
useful shorthand to this idea: ``if $x$ implies $y$, they have to be
compatible''.\cite{drieschner} However, to follow Jauch and Piron,
one has to make a substantial intuitive assumption: in order to use
the derivation of orthomodularity via the existence of Boolean
subalgebras, one must separately introduce a notion of compatibility
of propositions. It is clear that compatibility refers to the
concept of complementarity in quantum physics, and thus one uses an
intuitive feature motivated by quantum theory, yet before deriving
this theory and \textit{for} deriving it. Such an early recourse to
the quantum theoretic intuition is certainly an unwelcome factor.

In the present article we give a different conceptual justification
of orthomodularity of a lattice in the framework of the
information-theoretic approach to quantum
theory.\cite{WheIBM,FuchsInLight,BZ,grinbijqi} This approach
consists in viewing quantum theory as a theory about, or
\textit{of}, information. Constraints imposed on the kind of
information available in the theory lead then to a particular,
rather than general, theory of information, and one seeks for a set
of constraints which will produce quantum theory. To express the
constraints mathematically, a certain formalism must be employed,
e.g. quantum logical\cite{RovRQM,mydiss} or algebraic.\cite{Bub} In
the quantum logical formalism, we show that orthomodularity arises
as a consequence of the following information-theoretic
axiom:\begin{axiom} There is a maximum amount of relevant
information that can be extracted from a system.
\label{ax1}\end{axiom} \addtocounter{axiomcounter}{1} This axiom was
first formulated by Rovelli\cite{RovRQM} who took inspiration from
the work of Wheeler.\cite{whe89,whe92} Rovelli linked it to the idea
of maximal ``informational capacity'' of a system and argued
informally that Axiom~\ref{ax1} already adds the Planck constant to
physics. He, however, did not discuss the meaning, nor treat
rigorously, what we take to be the crucial term in the formulation
of the axiom: \textit{relevant} information. After introducing basic
concepts of the theory of orthomodular lattices in
Section~\ref{defs}, we propose in Section~\ref{sect2} a formal
definition of the notion of relevance of information. This leads to
a key result in Section~\ref{sect4} in which we prove that the
lattice of yes-no questions is orthomodular.

\bigskip\section{ELEMENTS OF LATTICE THEORY}\label{defs}

Lattice $\mathcal{L}$ is a partially ordered set in which any two
elements $x,y$ have a supremum $x\vee y$ and an infimum $x\wedge
y$. Equivalently, one can require that a set $\mathcal{L}$ be
equipped with two idempotent, commutative, and associative
operations $\vee,\wedge$ :
$\mathcal{L}\times\mathcal{L}\rightarrow \mathcal{L}$, which
satisfy $x\vee(y\wedge x)=x$ and $x\wedge (y\vee x)=x$. The
partial ordering is then defined by $x\leq y$ if $x\wedge y =x$.
The largest element in the lattice, if it exists, is denoted by
$1$, and the smallest one, if it exists, by $0$.

Lattice is called complete when every subset of $\mathcal{L}$ has
a supremum as well as an infimum. Complete lattice always contains
elements $0$ and $1$. An atom of lattice $\mathcal{L}$ is an
element $a$ for which $0\leq x\leq a$ implies that $x=0$ or $x=a$.
A lattice with $0$ is called atomic if for every $x\neq 0$ in
$\mathcal{L}$ there is an atom $a\neq 0$ such that $a\leq x$.

Lattice is called distributive if
\begin{equation}x\vee(y\wedge z)=(x\vee y)\wedge(x\vee z).\label{distrib}\end{equation}
One can weaken the distributivity condition by requiring
(\ref{distrib}) only if $x\leq z$. Thus, a lattice is called
modular if for all $y$ \begin{equation}x\leq z \Rightarrow x\vee
(y\wedge z)=(x\vee y)\wedge z.\label{modular}\end{equation}

Orthocomplemented lattice is a lattice $\mathcal{L}$ which carries
orthocomplementation, i.e. a map $x\mapsto x^{\bot}$, satisfying
for all $x,y\in \mathcal{L}$\label{oc} the following conditions:
  (a) $x^{\bot\bot}=x,$
  (b) $x\leq y \Leftrightarrow  y^{\bot}\leq x^{\bot},$
  (c) $x\wedge x^\bot =0,$
  (d) $x\vee x^\bot =1.$
In an orthocomplemented lattice hold de Morgan laws
\begin{equation} 1^\bot=0;\quad 0^\bot=1;\quad
(x\vee y)^\bot=x^\bot\wedge y^\bot;\quad (x\wedge
y)^\bot=x^\bot\vee y^\bot. \label{demorgan}\end{equation}

By further weakening modularity condition~(\ref{modular}), one
arrives at the following definition:

\begin{defn} Orthocomplemented lattice $\mathcal{L}$ is called
\textbf{orthomodular} if condition (\ref{modular}) holds for
$y=x^\bot$, that is,
\begin{equation}x\leq z \Rightarrow x\vee (x^\bot \wedge
z)=z.\label{orthomodular}
\end{equation}
\label{deforthomodular}\end{defn}

It is instructive to give the following reformulation of the
condition of orthomodularity,\cite{kalmbach83} for which we offer a
new proof.

\begin{lem}
An orthocomplemented lattice $\mathcal{L}$ is orthomodular if and
only if $x\leq z$ and $x^\bot \wedge z=0$ imply
$x=z$.\label{lemorthomod}
\end{lem}
\begin{proof}
If the lattice is orthomodular, i.e. (\ref{orthomodular}) holds,
and if $x^\bot \wedge z=0$, then $z=x \vee 0=x$. To prove the
converse, it suffices to show that if the lattice is not
orthomodular then there exist elements $x$ and $z$ such that
\begin{equation}
x\leq z,\quad x^\bot \wedge z=0,\quad x\neq z.\label{plusik1}
\end{equation}
Let us use the notation $x<z$ if $x\leq z$ and $x \neq z$. We can
then rewrite (\ref{plusik1}) as
\begin{equation}
x< z,\quad x^\bot \wedge z=0.\label{palka}
\end{equation}
Assume that the lattice is not orthomodular. In virtue of
Definition~\ref{deforthomodular}, there exist elements $y$ and $z$
such that
\begin{equation}
y\leq z,\quad y\vee(y^\bot \wedge z)\neq z.\label{goriz}
\end{equation}
Now recall that in any lattice holds\cite{cohn}
\begin{equation}
a\leq b\Rightarrow (c\wedge b)\vee a\leq (c\vee a)\wedge b
\quad\forall c. \label{multipl}
\end{equation}
Put in (\ref{multipl}) $a=y$, $b=z$, $c=y^\bot$. Follows that
\begin{equation}
(y^\bot \wedge z)\vee y\leq (y^\bot \vee y)\wedge
z.\label{multik1}
\end{equation}
In the right-hand side replace $y^\bot \vee y$ by $1$, and
$1\wedge z=z$. Equation (\ref{multik1}) then takes the form
\begin{equation}
(y^\bot \wedge z)\vee y\leq z.\label{multik2}
\end{equation}
From equations (\ref{multik2}) and (\ref{goriz}) one obtains that
\begin{equation}
(y^\bot \wedge z)\vee y< z.\label{multik3}
\end{equation}
On the other hand, from de Morgan laws (\ref{demorgan}) one has
\begin{eqnarray}
z\wedge(y\vee(y^\bot \wedge z))^\bot=z\wedge (y^\bot \wedge(y^\bot\wedge z)^\bot)=\nonumber\\
z\wedge (y^\bot\wedge(y\vee z^\bot))=(z\wedge y^\bot)\wedge(y\wedge z^\bot)=\nonumber\\
(z\wedge y^\bot)\wedge(z\wedge y^\bot)^\bot=0.\label{multik4}
\end{eqnarray}
Now put $x=y\vee(y^\bot \wedge z)$. Equations (\ref{multik3}) and
(\ref{multik4}) can be rewritten as
\begin{equation}
x<z,\quad x^\bot \wedge z=0.
\end{equation}
This is exactly what was required in (\ref{palka}).\end{proof}

\bigskip\section{INFORMATION-THEORETIC APPROACH}\label{sect2}

The fundamental notion of measurement in the information-theoretic
approach is represented as a yes-no question. Information is then
brought about in the answer to a yes-no question. Input of the
information-theoretic approach to the derivation of
orthomodularity is therefore limited to the choice of the
departure point: It is a set of yes-no questions that \textit{can
be asked} to the system. We postulate that this set is an
orthocomplemented complete atomic lattice $\BL$, with orthogonal
complementation denoting negation of a question.

At the level of ordinary linguistic usage of words, assume that
the information obtained from a question $a$ is \textit{relevant}
for the observer. We are looking for ways to make it irrelevant.
This can be achieved by asking some new question $b$ that will
make $a$ irrelevant. Consider, for instance, $b$ such that it
entails the negation of $a$: $b \rightarrow \neg a.$ If the
observer asks the question $a$ and obtains an answer to $a$, but
then asks a \textit{genuine} new question $b$, it means, by virtue
of what ``genuine'' commonly signifies, that the observer expects
either a positive or a negative answer to $b$. This, in turn, is
only possible if information $a$ is no more relevant; indeed,
otherwise the observer would be bound to always obtain the
negative answer to $b$. Consequently, we say that that, by asking
$b$, the observer makes the question $a$ irrelevant. Such
considerations of common linguistic usage of words motivate the
following formal definition of relevance.

\begin{defn}
Question $b$ is called \textbf{irrelevant} with respect to
question $a$ if $b\wedge a^\bot\neq 0$. Otherwise question $b$ is
called relevant with respect to question $a$.
\label{revd}\end{defn}

\begin{figure}[htbp]
\begin{center}
\epsfysize=2.5in \epsfbox{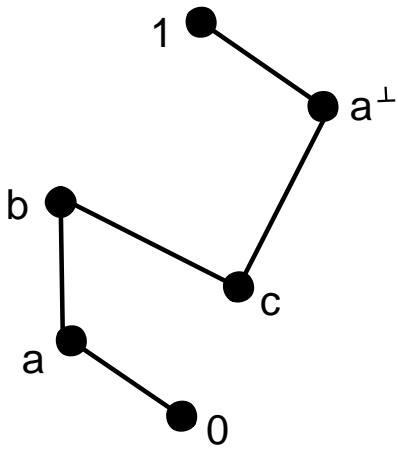}
\end{center}
\caption{\singlespacing Notion of relevance. Order in the lattice
is denoted by solid lines and grows from bottom to top, i.e.
$0\leq a\leq b$, etc. If there exists $c\neq 0$ such that $c\leq
b$ and $c\leq a^\bot$, then question $b$ is irrelevant with
respect to question $a$, i.e. in $b$ ``is contained a component''
of non-$a$, and consequently, by asking $b$, one renders
information of the question $a$ irrelevant.} \label{relevfig}
\end{figure}

What does Definition~\ref{revd} mean with regard to the best known
lattice, i.e. the Hilbert lattice or the lattice of all closed
subspaces of the Hilbert space? If $x\in \BL$ is a closed subspace
of the Hilbert space $H$, then its orthogonal complement
$x^\bot\in \BL$ satisfies $x\oplus x^\bot=H$ and $\dim x =
\mathrm{codim}\;x^\bot$. Assume, as on Figure~\ref{relevfig}, that
two Hilbert lattice elements $a$ and $b$ are such that $a < b$.
Then $\dim b > \dim a$ and, consequently, there exist closed
subspaces of $b$ not contained in $a$. Call one such subspace $c$.
In virtue of the definition of orthogonality, one has $a\oplus
a^\bot =H$, and this implies that $c\cap a^\bot\neq \emptyset.$
Therefore, $b\wedge a^\bot\neq 0$. We obtain that if $b>a$ in a
Hilbert lattice, then $b$ is always irrelevant with respect to
$a$. Generally, as follows from Lemma~\ref{lemorthomod} and
Definition~\ref{revd}, in an orthomodular lattice there exist no
questions relevant with respect to a given one, which at the same
time are strictly greater than this question.

Relevance then becomes trivially assimilated to orthogonality in
the case of Hilbert lattices. Any closed subspace $c$ such that $c
\subseteq a$ provides a lattice element $c$ relevant with respect
to $a$, and all other lattice elements are irrelevant with respect
to $a$. If, for example, one is concerned with the Hilbert space
of spin projections, with respect to $\sigma _x$ only operators
which commute with it are relevant, i.e. $\sigma _x$ itself and
the null projector. The interest in the notion of relevance is
therefore motivated, not by the Hilbert lattices, but by the
generic case where relevance is not a mere reformulation of
set-theoretic inclusion and of the compatibility of propositions.
While some theorists deliberately choose not to cross this limit,
the property of orthomodularity can be properly \textit{derived}
only if one goes beyond the standard Hilbert lattice vision.

\begin{figure}[htbp]
\begin{center}
\epsfysize=2.5in \epsfbox{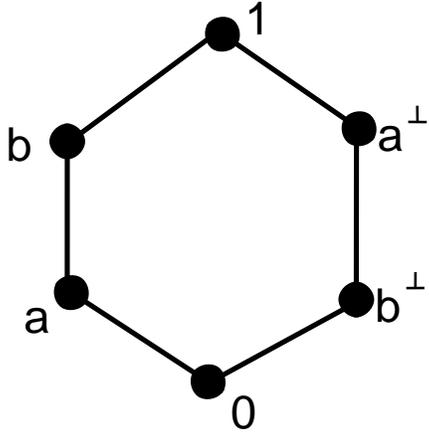}
\end{center}
\caption{\singlespacing An orthocomplemented non-orthomodular
lattice.}\label{nonortho}
\end{figure}

An example when the notion of relevance is non-trivial is given on
Figure~\ref{nonortho}. This is a non-orthomodular complete atomic
lattice, in which $b$ is relevant with respect to $a$, although
$b\geq a$ holds. In general, there exist infinitely many
non-orthomodular complete atomic orthocomplemented lattices, which
can be for example constructed by slightly violating Greechie's
procedure.\cite{greechie} All such lattices are ``wild'' from the
point of view of quantum theory, and it is necessary to give an
argument excluding them from consideration.

Apart from relevance, another concept mentioned in
Axiom~\ref{ax1}, which necessitates a formal definition, is the
amount of information. In general, when relevance of information
is not preserved throughout several acts of asking questions, i.e.
when, as new information comes in, some older information ceases
to be relevant, it is impossible to dissociate the amount of
information generated by a new question from the considerations of
relevance of this new question with respect to the previously
asked ones. Influence of every new question on the total amount of
relevant information is therefore contextual: it depends on which
questions have been asked before. It is easy, though, to define
one particular property of the amount of information in an
exceptional case when the contextual dependence disappears. Thus,
we assume that the amount of information is a monotonously growing
function of the number of answers to yes-no questions that the
observer obtains, in the exceptional case when none of the new
questions render irrelevant any of the previous questions. In
other words, if the observer keeps all the old information, then
the amount of information grows as new information comes in.

Going back to the phrase in the first paragraph in which we
introduced the set of yes-no questions, we interprete the words
``can be asked'' so that one is concerned with \textit{possible}
acts of bringing about information, i.e. the lattice contains all
possible questions, most of which, indeed, will never be asked.
Applied to the amount of information, such a lattice construction
leads to a further assumption that there always are sufficiently
many questions as to bring in, potentially, any amount of
information permitted \textit{a priori} by the theory. In
particular, there always exists a question such that it brings the
maximum amount of relevant information to be had about the system.

\bigskip\section{ORTHOMODULARITY OF A LATTICE}\label{sect4}

Axiom~\ref{ax1} ensures that cases as the one shown on
Figure~\ref{nonortho} do not appear in quantum theory, i.e. that
all important lattices are orthomodular. This is achieved in
virtue of the following theorem.

\begin{thm}
$\mathcal{L}$ is an orthomodular lattice.\label{orthothm}
\end{thm}

\begin{proof}
By Axiom~\ref{ax1} there exists a finite upper bound of the amount
of relevant information. Let this be an integer $N$. Select an
arbitrary question $a$ and consider a question $\tilde{a}$ such
that
\begin{equation}\{a,\tilde{a}\}\label{qqti}\end{equation} bring the maximum amount of
relevant information, i.e. $N$ bits. Notation $\{\ldots\}$ means a
sequence of questions that are asked one after another. Because
all information here is relevant, we have by the definition of
relevance that \begin{equation}\tilde a \wedge
a^\bot=0\label{qqti2}.\end{equation}

We now use Lemma~\ref{lemorthomod}. It is sufficient to show that
$a\leq b$ and $a^\bot\wedge b=0$ imply $a=b$. Note first that the
second condition means, by Definition~\ref{revd}, that $b$ is
relevant with respect to $a$. Since $a\leq b$, we obtain that
\begin{equation}b^\bot \leq a^\bot.\end{equation}
Using this result and the result of Equation~\ref{qqti2}, we
derive \begin{equation}\tilde a\wedge
b^\bot=0.\label{qqti3}\end{equation} This, in turn, means that
question $\tilde a$ is relevant with respect to $b$.

Now suppose, contrary to what is needed, that $b>a$ and consider
the following sequence of questions:
\begin{equation}\{a,\, b,\, \tilde a\}\end{equation}
From Equations~\ref{qqti2} and \ref{qqti3} follows that relevance
is not lost in this sequence of question, i.e. all later
information is relevant with respect to all earlier information.
However, while relevance is preserved, this sequence, in virtue of
the fact that $a\neq b$, brings about more information than
sequence (\ref{qqti}). It means that we have constructed a setting
in which the amount of relevant information is strictly greater
than $N$ bits, causing a contradiction with the initial
assumption. Consequently, $a=b$ and the lattice $\mathcal{L}$ is
orthomodular.
\end{proof}

\bigskip\section{CONCLUSION}

Theorem~\ref{orthothm} provides a solution to the long-standing
puzzle about the meaning of orthomodularity. Historically,
orthomodularity appeared at a late stage of quantum logical
research, the first invention being the theory of von Neumann
algebras,\cite{murvN36} second the theory of von Neumann modular
lattices,\cite{bvn} and only third the theory of orthomodular
lattices. Its origin can be traced back to Husimi's work in which he
was predominantly concerned with deriving non-Boolean logic from
empirical facts,\cite{husimi} and in full orthomodularity was for
the first time discussed much later.\cite{loomis} Thus, appearing as
the weakest lattice constraint and as a result of the historic
evolution of quantum logic, orthomodularity was easily taken for
granted in axiomatization attempts. Reviving recently the interest
in quantum logic, information-theoretic approach permits to bridge
this conceptual gap. From the information-theoretic viewpoint
orthomodularity must be seen as a consequence of the finite amount
of relevant information.

\bigskip\section*{ACKNOWLEDGEMENTS} The initial idea of proof of
Lemma~\ref{lemorthomod} is due to Prof.~V.~A.~Franke. Many thanks to
Carlo Rovelli and Jeffrey Bub for their insightful comments.

\bigskip

\begin{thebibliography}{99}

\bibitem{mackey63}
G.W. Mackey
\newblock {\em Mathematical Foundations of Quantum Mechanics}
\newblock (Benjamin, New York, 1963).

\bibitem{zieler}
N.~Zieler
\newblock ``Axioms for non-relativistic quantum mechanics.''
\newblock {\em Pacific J. Math.} \textbf{11}, 1151--1169 (1961).

\bibitem{varada2}
V.S. Varadarajan
\newblock {\em Geometry of Quantum Theory}
\newblock (Van Norstand, Princeton, 1968).

\bibitem{piron76}
C.~Piron
\newblock {\em Foundations of Quantum Physics}
\newblock (Benjamin, Reading, 1976).

\bibitem{kochspeck65}
S.~Kochen and E.P. Specker
\newblock ``Logical structures arising in quantum theory,''
\newblock in Addison J. et~al., eds., {\em The Theory of Models}
  (North-Holland, Amsterdam, 1965).

\bibitem{guenin}
J.~Guenin
\newblock ``Axiomatic formulations of quantum theories,''
\newblock {\em J. Math. Phys.} \textbf{7}, 271--282 (1966).

\bibitem{Gunson}
J.~Gunson
\newblock ``On the algebraic structure of quantum mechanics,''
\newblock {\em Comm. Math. Phys.} \textbf{6}, 262--285 (1967).

\bibitem{jauch}
J.M. Jauch
\newblock {\em Foundations of Quantum Mechanics}
\newblock (Addison-Wesley, Reading, 1968).

\bibitem{pool}
J.C.T. Pool
\newblock ``Baer $^{*}$-semigroups and the logic of quantum
mechanics,''
\newblock {\em Comm. Math. Phys.} \textbf{9}, 118--141 (1968).

\bibitem{plymen}
R.J. Plymen
\newblock ``A modification of {P}iron's axioms,''
\newblock {\em Helvetica Physica Acta} \textbf{41}, 69--74 (1968).

\bibitem{maczy}
M.J. Maczynski
\newblock ``On a functional representation of the lattice of projections on a
  {H}ilbert space,''
\newblock {\em Studia Mathematica} \textbf{47}, 253--259 (1973).

\bibitem{marlow}
A.R. Marlow
\newblock ``Orthomodular structures and physical theory,''
\newblock In A.R. Marlow, ed., {\em Mathematical Foundations of Quantum
  Theory} (Academic Press, New York, 1978).

\bibitem{holland}
S.S. Holland
\newblock ``Orthomodularity in infinite dimensions; a theorem of {M}.
  {S}ol\`{e}r,''
\newblock {\em Bull. Amer. Math. Soc.} \textbf{32}, 205--234 (1995).

\bibitem{ruttimann}
G.T. R\"uttimann
\newblock ``Projections on orthomodular lattices,''
\newblock In A.~Hartk\"amper and H.~Neumann, eds., {\em Foundations of
  Quantum Mechanics and Ordered Linear Spaces} (Springer, Berlin, 1974).

\bibitem{keller}
H.A. Keller
\newblock ``On the lattice of all closed subspaces of a hermitian
space,''
\newblock {\em Pacific J. Math.} \textbf{89}, 105--107 (1980).

\bibitem{bc}
E.G. Beltrametti and G.~Cassinelli
\newblock {\em The Logic of Quantum Mechanics}
\newblock (Addison-Wesley, Reading, 1981).

\bibitem{piron}
C.~Piron
\newblock ``Axiomatique quantique,''
\newblock {\em Helvetica Physica Acta} \textbf{36}, 439--468 (1964).

\bibitem{drieschner}
M.~Drieschner
\newblock ``Lattice theory, groups and space,''
\newblock In L.~Castell, M.~Drieschner, and C.F. von Weizs\"acker, eds.,
  {\em Quantum Theory and the Structures of Time and Space} (Carl
  Hansen, M\"unchen, 1975).

\bibitem{WheIBM}
J.A. Wheeler
\newblock ``World as system self-synthesized by quantum
networking,''
\newblock {\em IBM J. Res. Develop.} \textbf{32}, 4--15 (1988).

\bibitem{FuchsInLight}
C.A. Fuchs
\newblock ``Quantum foundations in the light of quantum
information,''
\newblock In A.~Gonis and P.E.A. Turchi, eds., {\em Decoherence and its
  Implications in Quantum Computation and Information Transfer} (IOS Press, Amsterdam, 2001).

\bibitem{BZ}
C.~Brukner and A.~Zeilinger
\newblock ``Information and fundamental elements of the structure of quantum
  theory,''
\newblock In L.~Castell and O.~Ischebeck, eds., {\em Time, Quantum,
  Information} (Springer, Berlin, 2003).

\bibitem{grinbijqi}
A.~Grinbaum
\newblock ``Elements of information-theoretic derivation of the formalism of
  quantum theory,''
\newblock {\em International Journal of Quantum Information} \textbf{1}, 289--300 (2003).

\bibitem{RovRQM}
C.~Rovelli
\newblock ``Relational quantum mechanics,''
\newblock {\em Int. J. of Theor. Phys.} \textbf{35}, 1637 (1996).

\bibitem{mydiss}
A.~Grinbaum
\newblock {\em The Significance of Information in Quantum Theory}
\newblock (PhD thesis, Ecole Polytechnique, Paris, 2004).

\bibitem{Bub}
R.~Clifton, J.~Bub, and H.~Halvorson
\newblock ``Characterizing quantum theory in terms of information-theoretic
  constraints,''
\newblock {\em Found. Phys.}, \textbf{33}, 1561--1591 (2003).

\bibitem{whe89}
J.A. Wheeler
\newblock ``Information, physics, quantum: the search for the
links,''
\newblock In {\em Proceedings of 3rd International Symposium Foundations of
  Quantum Mechanics} (Tokyo, 1989).

\bibitem{whe92}
J.A. Wheeler
\newblock ``It from bit,''
\newblock In L.~Keldysh and V.~Feinberg, eds., {\em Sakharov Memorial
  Lectures on Physics, Vol.~2} (Nova Science, New York, 1992).

\bibitem{kalmbach83}
G.~Kalmbach
\newblock {\em Orthomodular Lattices}
\newblock (Academic Press, London, 1983).

\bibitem{cohn}
P.M. Cohn
\newblock {\em Universal Algebra}
\newblock (Harper and Row, New York, 1965).

\bibitem{greechie}
R.J. Greechie
\newblock ``Orthomodular lattices admitting no states,''
\newblock {\em Journal of Combinatorial Theory A} \textbf{10}, 119--132 (1971).

\bibitem{murvN36}
F.J. Murray and J.~{von Neumann}
\newblock ``On rings of operators,''
\newblock {\em Ann. of Math.} \textbf{37}, 116--229 (1936).

\bibitem{bvn}
G.~Birkhoff and J.~{von Neumann}
\newblock ``The logic of quantum mechanics,''
\newblock {\em Ann. Math. Phys.} \textbf{37}, 823--843 (1936).

\bibitem{husimi}
K.~Husimi
\newblock ``Studies in the foundations of quantum mechanics,''
\newblock {\em Proceedings of the Physico-Mathematical Society of Japan}
  \textbf{19}, 766--789 (1937).

\bibitem{loomis}
L.H. Loomis
\newblock {\em The Lattice Theoretic Background of the Dimension Theory of
  Operator Algebras}
\newblock (American Mathematical Society, Providence, 1955).

\end{thebibliography}

\end{document}